\documentclass[11pt,a4paper,english]{article}
\usepackage{float}
\usepackage{graphicx}
\usepackage{amsmath}
\usepackage{amsfonts}
\usepackage{latexsym}

\usepackage[dvips]{hyperref}

\newcommand{\beq}{\begin{equation}}
\newcommand{\eeq}{\end{equation}}

\newcommand{\intd}{\mathrm{d}}

\newcommand{\bra}[1]{\langle #1 |}
\newcommand{\ket}[1]{|#1 \rangle}
\newcommand{\braket}[2]{\langle #1 | #2 \rangle}
\newcommand{\brakett}[1]{\langle #1 \rangle}
\newcommand{\brakettt}[3]{\langle #1 | #2 |#3 \rangle}

\newcommand{\iin}{\textit{in} }
\newcommand{\oout}{\textit{out} }
\newcommand{\kk}{\mathrm{k}}
\newcommand{\xx}{\mathrm{x}}
\newcommand{\ii}{\mathrm{i}}
\newcommand{\ren}{\mathrm{as}}

\usepackage{babel}

\begin{document}

\begin{center}
{\LARGE
On some aspects of the definition of scattering states\\[2mm] in quantum field theory
}
\vspace{1.5cm}

{\large
G\'{a}bor Zsolt T\'{o}th
\vspace{1cm}
}

\textit{
Research Institute for Particle and Nuclear Physics, \\
Hungarian Academy of
Sciences, P.O.B.\ 49, 1525 Budapest, Hungary}\\[8mm]
e-mail:\ \  tgzs@rmki.kfki.hu 

\end{center}
\vspace{0.1cm}

\begin{center}
{\bf Abstract}\\[8mm]

\begin{tabular}{p{14cm}}
{\small
The problem of extending quantum-mechanical formal scattering theory to a more general class of models that also includes quantum field theories is discussed, 
with the aim of clarifying certain aspects of the definition of scattering states. 
As the strong limit is not suitable for the definition of scattering states in quantum field theory, 
some other limiting procedure is needed. Two possibilities are considered, the abelian limit and
adiabatic switching. 
Formulas for the scattering states based on both methods are discussed, and it is found that generally 
there are significant differences between the two approaches. 
As an illustration of the application and the features of these formulas, S-matrix elements and energy corrections in two quantum field theoretical models are calculated using (generalized) old-fashioned perturbation theory. The two methods are found to give equivalent results.
}
\end{tabular}
\end{center}

\vspace{0.8cm}
\noindent
\hspace{1cm}PACS codes: 11.80.-m; 03.70.+k;  03.65.Nk

\thispagestyle{empty}

\newpage

\section{Introduction}

Much of our knowledge about atomic and subatomic physics comes from studying the results of scattering experiments, 
therefore it is not surprising that special attention has been devoted to the theoretical investigation of scattering processes.  
The central objects in the quantum theoretical description of scattering processes are the S-matrix elements, which are scalar products of scattering states, also known as  \iin and \oout states:
\begin{equation}
\label{eq.1}
S_{wv}=\braket{w,out}{v,in}.
\end{equation} 
$\ket{v,in}$ is a state that looks, in a suitable sense, like the 
free-particle 
state $\ket{v}$ in the remote past, and $\ket{w,out}$ is a state that looks like the free-particle state 
$\ket{w}$ in the far future (see e.g.\ chapter 3 of \cite{W}).  
A central issue in scattering theory is to define properly these \iin and \oout states.
In quantum-mechanical potential scattering,  standard formulas (see \cite{W,Taylor,Newton}) that  define the  \iin and \oout states 
are the following:
\begin{eqnarray}
\label{eq.2in}
\ket{v,in} & = & \lim_{T\to\infty} 
U(0,-T)\ket{v},\\
\label{eq.2out}
\ket{w,out} & = & \lim_{T\to\infty} 
U(0,T)\ket{w},
\end{eqnarray}
where $U(t_2,t_1)$ is the time-evolution operator
\beq
U(t_2,t_1)=e^{\ii H_0t_2}e^{-\ii H(t_2-t_1)}e^{-\ii H_0t_1},
\eeq 
$H$ is the  total Hamiltonian operator that describes the scattering process, 
$\ket{v}$ and $\ket{w}$ are state vectors that characterize the scattering particles in the infinite past and future, respectively, 
and $H_0$ is a free-particle Hamiltonian operator that describes the time evolution of  $\ket{v}$ and $\ket{w}$.

It is an interesting question whether it is possible to generalize (\ref{eq.2in}) and  (\ref{eq.2out}) for a wider class of models, including
quantum field theoretical models in particular. 
It is well known that in relativistic quantum field theory there exists a standard covariant formalism, based on the 
Lehmann-Symanzik-Zimmermann (LSZ) reduction formulas, for the description of scattering. This formalism, which is presented in several textbooks,
is specific to field theory in the sense that it involves the fields that appear in the model to which it is applied.
In the present paper 
we are interested in generalizations of (\ref{eq.2in}) and  (\ref{eq.2out}) that maintain the feature of only involving 
Hamiltonian operators and state vectors, but no further details (e.g.\ symmetry properties) 
of the structure of the model
to which they are applied. 
Besides their theoretical interest,
such generalizations of (\ref{eq.2in}) and  (\ref{eq.2out}) can also be useful as starting points for old-fashioned 
perturbation theory, which has the distinguished 
property of being formulated solely on the basis of 
on-shell particles, and is thus more suitable for certain purposes than the covariant perturbation theory.  
Moreover, in  $1+1$ dimensional QFT 
there exist models for which it is not known how to apply the standard formalism \cite{DMS2}.

The problem of generalizing (\ref{eq.2in}) and  (\ref{eq.2out}) in a not specifically field theoretical way 
is addressed in many quantum field theory textbooks in preparation for the
presentation of the field theoretical formalism 
(see e.g.\ chapter 3 of \cite{W}). 
However, it seems to us that the literature on this subject is incomplete, in particular regarding the normalization 
of the \iin and \oout states and the treatment of the $T\to\infty$ limit.
Correspondingly, we feel that some details of the (generalized) old-fashioned perturbation theory should also be revisited.

The aim of this paper is 
to discuss these aspects in two significantly different 
generalizations of (\ref{eq.2in}) and  (\ref{eq.2out}).
In section \ref{sec.ab}, we consider the case in which the abelian average is used in the definition of the \iin and \oout states.
In section \ref{sec.ad}, we discuss another generalization of (\ref{eq.2in}) and  (\ref{eq.2out}), which we proposed recently in \cite{T}, and which is  
based on an adiabatic switching. 
In section \ref{sec.ex}, we apply the formulas presented in 
sections \ref{sec.ab} and \ref{sec.ad} 
to two quantum field theoretical models, in the framework of (generalized) old-fashioned perturbation theory. 
The first model describes the scattering of a massive scalar particle on a fixed defect line in $1+1$ dimensions. 
The second example is the $\phi^4$ model. This section is intended to provide an illustration for sections  \ref{sec.ab} and \ref{sec.ad}.
Our conclusions are given in section \ref{sec.c}. 

Although the problems discussed in sections \ref{sec.ab} and \ref{sec.ad} are somewhat mathematical in nature, 
we try to avoid too much mathematical sophistication;  
nevertheless we mention that the limits of state vectors in Hilbert spaces are always understood to be strong limits.

\section{Abelian limit}
\label{sec.ab}

The first generalization of (\ref{eq.2in}) and (\ref{eq.2out})  that we consider is the following: 
\begin{eqnarray}
\label{eq.a3}
\ket{v,in} & = & \frac{1}{\sqrt{Z_{v,in}}}\lim_{\epsilon\to +0} \lim_{T\to\infty} \hat{U}_\epsilon (0,-T) \ket{v},\\
\label{eq.a4}
\ket{w,out} & = & \frac{1}{\sqrt{Z_{w,out}}}\lim_{\epsilon\to +0} \lim_{T\to\infty} \hat{U}_\epsilon (0,T) \ket{w},
\end{eqnarray}
where  
\begin{eqnarray}
\label{eq.a5}
Z_{v,in} & = & \lim_{\epsilon\to +0} \lim_{T\to\infty}
\frac{\brakettt{v}{\hat{U}_\epsilon(0,-T)^\dagger\hat{U}_\epsilon (0,-T)}{v}}{\braket{v}{v}},\\
\label{eq.a6}
Z_{w,out} & = & \lim_{\epsilon\to +0} \lim_{T\to\infty}
\frac{\brakettt{w}{\hat{U}_\epsilon(0,T)^\dagger\hat{U}_\epsilon (0,T)}{w}}{\braket{w}{w}},
\end{eqnarray}
\begin{eqnarray}
\label{eq.a7}
\hat{U}_\epsilon (0,-T) & = & \epsilon\int_0^T \intd \tau\, e^{-\epsilon \tau}\bar{U} (0,-\tau),\\
\label{eq.a8}
\hat{U}_\epsilon (0,T) & = & \epsilon\int_0^T \intd \tau\, e^{-\epsilon \tau}\bar{U} (0,\tau),
\end{eqnarray}
$\epsilon$ is a small positive real number, and 
\beq
\label{eq.a9}
\bar{U} (t_2,t_1)=e^{\ii H^\ren t_2}e^{-\ii H(t_2-t_1)}e^{-\ii H^\ren t_1}.
\eeq 

Three important differences between (\ref{eq.a3}), (\ref{eq.a4}) and (\ref{eq.2in}), (\ref{eq.2out}) should be mentioned. 
The first one is that $H^\ren$ is written instead of $H_0$.   
$H^\ren$ is a suitably chosen Hamiltonian operator that describes the time evolution of the scattering particles at times long before and after
the scattering event
and thus has a role similar to that of $H_0$. The purpose of introducing the notation $H^\ren$ is to emphasize the fact that 
due to self-interaction effects 
$H^\ren$ is generally not 
identical to the Hamiltonian obtained from $H$ by switching off the interaction (see e.g.\ section 3.1 of \cite{W}), 
in contrast with the situation in potential scattering. The superscript ${}^\ren$ is intended to refer to the word {\sl asymptotic}. 
Sometimes $H^\ren$ is also called renormalized $H_0$ operator ($H_0$ being now the
Hamiltonian obtained from $H$ by switching off the interaction). 
We also note that in quantum field theory it is usual to modify $H$ rather than $H_0$ (see the second example in section \ref{sec.ex}). 

The second difference is the replacement of the simple $T\to\pm\infty$ limits by abelian limits 
(i.e., using $\lim_{\epsilon\to +0} \lim_{T\to\infty} \hat{U}_\epsilon (0,-T) \ket{v}$ 
and $\lim_{\epsilon\to +0} \lim_{T\to\infty} \hat{U}_\epsilon (0,T) \ket{w}$
instead of\\ $\lim_{T\to\infty} \bar{U} (0,-T) \ket{v}$ and 
$\lim_{T\to\infty} \bar{U} (0,T) \ket{w}$. 
Nevertheless, both the $T\to\infty$ and the $\epsilon\to +0$ limits are assumed to be strong limits.). 
The abelian limit is frequently applied in the literature,
e.g.\ in \cite{GG,GW}, and in chapter 3 of \cite{W} (in the latter reference the term ``abelian limit" is not used).
It is often not mentioned in the literature, but it should be emphasized that, as explained in more detail below, 
the simple limits $\lim_{T\to\infty} \bar{U} (0,-T) \ket{v}$ etc.\ are not sufficient in general for quantum field theories, as they usually do not exist.
In particular, this situation arises in the examples presented in section \ref{sec.ex}.

The third main difference is the presence of the factors $\frac{1}{\sqrt{Z_{v,in}}}$ and 
$\frac{1}{\sqrt{Z_{w,out}}}$, which are included in order to ensure the correct normalization of $\ket{v,in}$ and 
$\ket{w,out}$. The need for such normalization factors is related to the use of the abelian limit, as explained below under 1.) and 2.) in more detail.
Although the need for these factors 
was recognized in some of the early literature on scattering in quantum field theory (see 
section 5.7 of \cite{GW}), the reasons are not explained in much detail.   
In more recent books and articles these normalization factors do not appear.
The authors of \cite{PS} are concerned with the normalization of the \iin and \oout states 
(see sections 4.1 and 4.5 of \cite{PS}), but it is not the abelian limit that they apply, and the discussion is 
restricted mainly to the vacuum state.\footnote{It is important to note that 
the normalization of the \iin and \oout states  and the $T\to\infty$ limit are satisfactorily treated in those parts of the literature where the standard covariant field theoretical formalism and the LSZ reduction formulas are presented.}

In the remainder of this section, the discussion of the features of (\ref{eq.a3}) and (\ref{eq.a4}) is continued, but 
in order to separate clearly the various comments they are presented in the form of a numbered list.

\vspace{3mm}
1.) {\sl Relation between the abelian and simple limits}\\
It is well known that the abelian limit is more effective than the simple limit; in particular if 
$\lim_{T\to \infty} U(0,-T)\ket{v} = \ket{V}$ exists, where $U(0,-T)$ is a unitary operator for any value of $T$ and $\ket{v}$ 
is a vector of unit norm, then the abelian limit 
$\lim_{\epsilon\to +0}\epsilon \int_0^\infty \intd\tau\, e^{-\epsilon\tau} U(0,-\tau)\ket{v}$ also exists and is equal to $\ket{V}$. 
We now recall the proof of this result, which can be found in ,e.g., \cite{Taylor}. 
The integral $\int_0^\infty \intd \tau\, e^{-\epsilon \tau} U(0,-\tau)\ket{v}$ is convergent for any $\epsilon>0$, 
since  $\int_0^\infty e^{-\epsilon \tau} ||U(0,-\tau)\ket{v}||= \int_0^\infty \intd \tau\, e^{-\epsilon \tau} <\infty$, 
where $||U(0,-\tau)\ket{v}||=||v||=1$ has been used, which follows from the unitarity of $U(0,-\tau)$.
We define $\ket{V_\epsilon}$ as $\ket{V_\epsilon}= \epsilon \int_0^\infty \intd \tau\, e^{-\epsilon \tau} U(0,-\tau)\ket{v}$. 
It has to be shown that $\lim_{\epsilon\to +0}\ket{V_\epsilon}=\ket{V}$, i.e.\ 
$\lim_{\epsilon\to +0} ||V_\epsilon-V||=0$. This can be done as follows: let us split the integral 
$\ket{V_\epsilon} -\ket{V}=\epsilon \int_0^\infty \intd \tau\, e^{-\epsilon \tau} [U(0,-\tau)\ket{v}-\ket{V}]$ into two parts as
$\epsilon \int_0^{T_1} \intd \tau\, e^{-\epsilon \tau} [U(0,-\tau)\ket{v}-\ket{V}]+
\epsilon \int_{T_1}^\infty \intd \tau\, e^{-\epsilon \tau} [U(0,-\tau)\ket{v}-\ket{V}]
$. The norm of the first term goes to zero in the $\epsilon\to +0$ limit for any fixed $T_1>0$ because of the 
factor $\epsilon$. The $\epsilon\to +0$ limit of the norm of the second term can be estimated as 
$\sup_{\tau \in [T_1,\infty)} ||U(0,-\tau)\ket{v}-\ket{V} ||$, which can be made arbitrarily close to zero by choosing $T_1$ sufficiently large because of the convergence of $U(0,-\tau)\ket{v}-\ket{V}$ to zero as $\tau\to\infty$.   

\vspace{3mm}
2.) {\sl The reason for the normalization factors}\\
The factors $\frac{1}{\sqrt{Z_{v,in}}}$ and $\frac{1}{\sqrt{Z_{w,out}}}$ are included in (\ref{eq.a3}), (\ref{eq.a4}) in order to ensure that 
$\ket{v,in}$ and $\ket{w,out}$ have the same norm as $\ket{v}$ and $\ket{w}$.
The result above under point 1.) shows that if the simple limit $\lim_{T\to\infty} \bar{U} (0,-T) \ket{v}$ exists, then $Z_{v,in}=1$, since in this case the abelian limit   
$\lim_{\epsilon\to +0}\lim_{T\to\infty} \hat{U}_\epsilon (0,-T) \ket{v}$ is the same as the simple limit $\lim_{T\to\infty} \bar{U} (0,-T) \ket{v}$, 
and the norm of the latter is the same as the norm of $\ket{v}$ because of the unitarity of $\bar{U} (0,-T)$. 
However, the abelian limit can exist even if the simple limit does not, 
and in this case the norm of the vector obtained by the abelian limit can be different from the norm of $\ket{v}$, 
i.e.\ $Z_{v,in}\ne 1$ is possible. The same applies to the \oout states as well.

It is not only the normalization of the \iin and \oout states that may be affected when the abelian limit is used, but also their 
orthogonality properties.   
If $\ket{v_1}$ and $\ket{v_2}$ are orthogonal and the limits $\lim_{T\to\infty} \bar{U} (0,-T) \ket{v_1}$ 
and  $\lim_{T\to\infty} \bar{U} (0,-T) \ket{v_2}$ exist, then  $\ket{v_1,in}$ and $\ket{v_2,in}$ 
are also orthogonal because of the unitarity of $\bar{U} (0,-T)$. 
In general, however, the orthogonality of 
$\ket{v_1}$ and $\ket{v_2}$ does not imply the orthogonality of $\ket{v_1,in}$ and $\ket{v_2,in}$.
Nevertheless, the S-matrix is often defined as 
$S_{ij}=\braket{v_i,out}{v_j,in}$, where $\ket{v_j,in}$ and $\ket{v_i,out}$ are produced from a specially chosen set of orthonormal state vectors 
$\{ \ket{v_i} \}$  of the 
Hilbert space. These vectors usually represent plane waves (with the exception of the vacuum state); 
the index $i$ is a general multi-index and orthonormality means 
$\braket{v_i}{v_j}=\delta(i,j)$, where $\delta(i,j)$ is a Dirac-delta. 
$\ket{v_i}$ should be regarded as a vector-valued distribution in the variable $i$, 
and $S_{ij}$ is also a generalized function of $i$ and $j$. For the unitarity of $S_{ij}$ 
it is sufficient that $\ket{v_i,in}$ and $\ket{v_j,in}$ (and also $\ket{v_i,out}$ and $\ket{v_j,out}$)
be orthogonal if $i\ne j$, and the Hilbert spaces spanned by $\{ \ket{v_i,in} \}$ be the same as that spanned by $\{ \ket{v_i,out} \}$. 
In principle, it has to be verified in each particular case that these conditions are satisfied. 
The verification of orthogonality may be simplified by symmetries.

It seems that in the literature 
it is usually assumed that limits like 
$\lim_{T\to\infty} \bar{U} (0,-T) \ket{v}$ etc.\ exist, in which case the normalization factors $Z_{v,in}$ and $Z_{w,out}$ are equal to $1$.
However, in section \ref{sec.ex} and in point 4.) below we present examples showing the relevance of these factors; i.e.\ values 
$Z_{v,in}\ne 1$ and $Z_{w,out}\ne 1$
will be obtained in these examples, which also implies the nonexistence of the simple limits taken without the abelian averaging.

If the \iin and \oout states are produced from plane-wave states in  (\ref{eq.a3}) and (\ref{eq.a4}), then the evaluation of (\ref{eq.a5}) and (\ref{eq.a6}) 
requires some consideration, since plane wave states do not have finite norm.

\vspace{3mm}
3.) {\sl Intertwining property of the mappings defined by (\ref{eq.a3}) and (\ref{eq.a4}) }\\
Energy conservation in scattering processes is expressed by the property that
the mappings $\ket{v}\to\ket{v,in}$ and $\ket{w}\to \ket{w,out}$ defined by (\ref{eq.a3}) and (\ref{eq.a4}) intertwine $e^{\ii H^\ren \Delta t}$ and 
$e^{\ii H\Delta t}$, i.e.\
$\ket{e^{\ii H^\ren\Delta t} v,in}=e^{\ii H\Delta t}\ket{v,in}$ and $\ket{e^{\ii H^\ren\Delta t} w,out}=e^{\ii H\Delta t}\ket{w,out}$, where $\Delta t$ is an arbitrary positive real number. In order to derive these equations, let us consider  
$\ket{e^{\ii H^\ren\Delta t}v,in}$. We have 
\beq
\nonumber
\epsilon \int_0^\infty \intd \tau\, e^{-\epsilon \tau} \bar{U}(0,-\tau) e^{\ii H^\ren\Delta t}\ket{v}=
e^{\ii H\Delta t} e^{\epsilon\Delta t}\epsilon \int_0^\infty \intd \tau\,  e^{-\epsilon(\tau+\Delta t)} \bar{U}(0,-(\tau+\Delta t))\ket{v}
\eeq
\beq
\nonumber
=e^{\ii H\Delta t} e^{\epsilon\Delta t} \epsilon \int_{\Delta t}^\infty \intd \tilde{\tau}\,  e^{-\epsilon\tilde{\tau}} \bar{U}(0,-\tilde{\tau})\ket{v}
\eeq
\beq
\label{eq.14}
=e^{\ii H\Delta t} e^{\epsilon\Delta t} \epsilon \int_{0}^\infty \intd \tilde{\tau}\,  e^{-\epsilon\tilde{\tau}} \bar{U}(0,-\tilde{\tau})\ket{v}
-e^{\ii H\Delta t} e^{\epsilon\Delta t} \epsilon \int_{0}^{\Delta t} \intd \tilde{\tau}\,  e^{-\epsilon\tilde{\tau}} \bar{U}(0,-\tilde{\tau})\ket{v}.
\eeq
If $\epsilon\to 0$, then the second term in (\ref{eq.14}) obviously tends to zero and $e^{\epsilon\Delta t}$ tends to $1$. Thus, we have
\beq
\label{eq.15}
\lim_{\epsilon\to+0}\epsilon \int_0^\infty \intd \tau\, e^{-\epsilon \tau} \bar{U}(0,-\tau) \ket{e^{\ii H^\ren\Delta t} v}=
e^{\ii H\Delta t}  \lim_{\epsilon\to+0}  \epsilon \int_{0}^\infty \intd \tau\,  e^{-\epsilon\tau} \bar{U}(0,-\tau)\ket{v}.
\eeq
Equation (\ref{eq.15}) shows that $Z_{v,in}=Z_{e^{\ii H^\ren\Delta t}v,in}$, thus dividing (\ref{eq.15}) by $\sqrt{Z_{v,in}}$ yields the desired result
$\ket{e^{\ii H^\ren\Delta t} v,in}=e^{\ii H\Delta t}\ket{v,in}$. 

Differentiating (\ref{eq.15}) with respect to $\Delta t$ and then setting $\Delta t=0$ also yields the intertwining relation
\beq
\lim_{\epsilon\to+0}\epsilon \int_0^\infty \intd \tau\, e^{-\epsilon \tau} \bar{U}(0,-\tau) \ket{H^\ren v}=
H  \lim_{\epsilon\to+0}  \epsilon \int_{0}^\infty \intd \tau\,  e^{-\epsilon\tau} \bar{U}(0,-\tau)\ket{v},
\eeq
i.e.\ 
$\lim_{\epsilon\to +0} \lim_{T\to\infty} \hat{U}_\epsilon (0,-T) \ket{H^\ren v}=
H \lim_{\epsilon\to +0} \lim_{T\to\infty} \hat{U}_\epsilon (0,-T) \ket{v}$. This implies that if $\ket{v}$ is an eigenvector of $H^\ren$, then $\ket{v,in}$ is an eigenvector of $H$ with the same eigenvalue. The case of \oout states is similar.

\vspace{3mm}
4.) {\sl The case of finite-dimensional Hilbert spaces}\\
Let us now consider the case of a Hilbert space with finite dimension $N$. 
Although there is no nontrivial scattering in this case, 
it is worth discussing because of its mathematical simplicity, and because it has similarities 
to more complicated cases (e.g.\ to those studied in section \ref{sec.ex}).
It also shows that the three features of (\ref{eq.a3}) and (\ref{eq.a4}) mentioned above are not specific to infinite-dimensional Hilbert spaces or to
quantum field theory.

It is straightforward to verify that the abelian limit $\lim_{\epsilon\to +0} \lim_{T\to\infty} \hat{U}_\epsilon (0,-T) \ket{v}=\ket{V}$ exists for any $\ket{v}$ vector and any $H^\ren$, and if  $\ket{v}$ is an eigenvector of  $H^\ren$ with eigenvalue $E_v$, then $\ket{V}$ is the orthogonal projection of $\ket{v}$ on the eigenspace of $H$ belonging to the eigenvalue $E_v$. In particular, $\ket{V}=0$ if $E_v$ is not an eigenvalue of $H$, 
which shows that $H^\ren$ should be chosen 
in such a way that it has the same eigenvalues as $H$. Similar statements can be made if $\hat{U}_\epsilon (0,T)$ is taken instead of 
$\hat{U}_\epsilon (0,-T)$. Moreover, the equation 
$\lim_{\epsilon\to +0} \lim_{T\to\infty} \hat{U}_\epsilon (0,-T) \ket{v} =
\lim_{\epsilon\to +0} \lim_{T\to\infty} \hat{U}_\epsilon (0,T) \ket{v}$ and thus $Z_{v,in}=Z_{v,out}$ 
also hold.

Let $\ket{v}$ be an eigenvector of $H^\ren$. In this case, $Z_{v,in}=1$ if and only if $\ket{v}$ is also an eigenvector of 
$H$ and in addition $H$ and $H^\ren$ have the same eigenvalue on $\ket{v}$; otherwise $Z_{v,in}<1$. A similar statement can be made for 
$Z_{v,out}$.
 
Assume now that the eigenvalues of $H$ are nondegenerate and $H^\ren$ has the same eigenvalues as $H$, 
and let $\ket{v_i^0}$, $i=1\dots N$ be a complete set of normalized eigenstates of $H^\ren$. 
Also assuming that $\ket{v_i^0}$ has nonzero projection on the eigenstate of $H$ that belongs to $E_{v_i}$ for all $i=1\dots N$, i.e.\ 
$Z_{v_i,in}=Z_{v_i,out}\ne 0$, 
the S-matrix $S_{ij}=\braket{v_i,out}{v_j,in}$ turns out to be the unit matrix ($\delta_{ij}$), as one expects. 
We note that the abelian limit allows a much larger class of suitable  $H^\ren$ operators than the simple limit; if $\ket{v}$ is an eigenvector of $H^\ren$, 
then $\bar{U} (0,-T)\ket{v}$ is an oscillating function of $T$ (and hence not convergent as $T\to\infty$) unless $\ket{v}$ is also an eigenvector of $H$ with the same eigenvalue. This means that if one demands that $\lim_{T\to \infty}\bar{U} (0,-T)\ket{v}$ should exist for any vector $\ket{v}$, 
then $H^\ren=H$ is necessary.
On the other hand, if the abelian limit is used, then any $H^\ren$ is suitable that has the same eigenvalues as $H$ and has the property 
that any eigenvector of it has nonzero projection on the eigenvector of $H$ with the same eigenvalue.

Let us now allow for $H$ to have degenerate eigenvalues, and let $H^\ren$ have the same eigenvalues as $H$ with the same degeneracies. Let $E_i$ and $E_j$ 
be two eigenvalues of $H$, and let $\ket{v_i}$ and $\ket{v_j}$ be two orthogonal and normalized eigenvectors of $H^\ren$ with the respective eigenvalues $E_i$ and $E_j$. 
If $E_i\ne E_j$, then $\braket{v_i,in}{v_j,in}=0$, of course. However, if $E_i = E_j$, then $\braket{v_i,in}{v_j,in}=0$
does not follow from $\braket{v_i}{v_j}=0$. Nevertheless, it is always possible to choose the (orthonormal) basis vectors within an eigenspace of $H^\ren$ in such a way that the corresponding \iin states are also orthonormal.
This can be seen in the following way: let $E$ be an eigenvalue and $V_E$ the corresponding eigenspace of $H^\ren$.
A second scalar product on $V_E$ can be defined as $\braket{u}{v}_2=\braket{u,in}{v,in}\sqrt{Z_{u,in}Z_{v,in}}$. This determines a self-adjoint linear map $A$ on $V_E$ with the property 
$\braket{u}{Av}=\braket{u}{v}_2$. 
$A$ admits a diagonal eigenbasis ${\ket{v_i}}$; for these basis vectors $\braket{v_i,in}{v_j,in}=\delta_{ij}$ and $\braket{v_i,out}{v_j,out}=\delta_{ij}$ hold.
In addition, $\braket{v_i,out}{v_j,in}=\delta_{ij}$ also holds.  
Furthermore, if there is a symmetry group whose action on the Hilbert space commutes with both $H$ and $H^\ren$, then $A$ is an invariant mapping, 
therefore if the representation of the symmetry group on $V_E$ is irreducible, then $A$ is the identity map, and thus $\braket{v_i,in}{v_j,in}=\delta_{ij}$ and
$\braket{v_i,out}{v_j,out}=\delta_{ij}$ (as well as $\braket{v_i,out}{v_j,in}=\delta_{ij}$)
hold for any orthonormal basis of $V_E$.

\vspace{3mm}
5.) {\sl Perturbation theory}\\
Let $\{\ket{v_i}\}$ be an orthonormal basis (i.e.\ $\braket{v_i}{v_j}=\delta(i,j)$) consisting of eigenvectors of $H^\ren$. 
The eigenvalue of $H^\ren$ on 
$\ket{v_i}$ is denoted by $E_i$.  

A perturbation series for $\lim_{T\to\infty}\brakettt{v_i}{\hat{U}_\epsilon (0,-T)}{v_j}$ 
can be obtained from the Dyson series; it reads as
\begin{multline}
\label{eq.t1}
\lim_{T\to
  \infty}\brakettt{v_i}{\hat{U}_\epsilon(0,-T)}{v_j}=
\braket{v_i}{v_j} -\ii  \frac{\brakett{i j}}{P(i j)+\epsilon}\\
+\sum_{k=2}^\infty (-\ii)^k 
\int \intd m_1 \intd m_2 \dots \intd m_{k-1}
\frac{\brakett{i m_{k-1}}}{P(i j)+\epsilon}
\frac{\brakett{m_{k-1} m_{k-2}}}{P(m_{k-1} j)+\epsilon}
\dots
\frac{\brakett{m_2 m_1}}{P(m_2 j)+\epsilon}
\frac{\brakett{m_1 j}}{P(m_1 j)+\epsilon},
\end{multline}
where the notation
\beq
\label{eq.not}
P(ij)=\ii (E_i-E_j),\qquad \brakett{ij}= 
\brakettt{v_i}{H-H^\ren}{v_j}
\eeq
is used. Similar formulas can be found e.g.\ in chapter 3 of \cite{W} and in \cite{DMS2}.

For  $\lim_{T\to\infty}\brakettt{v_i}{\hat{U}_\epsilon (0,T)^\dagger}{v_j}$ we have 
\begin{multline}
\label{eq.t12}
\lim_{T\to
  \infty}\brakettt{v_i}{\hat{U}_\epsilon(0,T)^\dagger}{v_j}=
\braket{v_i}{v_j} -\ii  \frac{\brakett{i j}}{P(ji)+\epsilon}\\
+\sum_{k=2}^\infty \left(-\ii\right)^k 
\int \intd m_1 \intd m_2 \dots \intd m_{k-1}
\frac{\brakett{i m_{1}}}{P(m_1 i)+\epsilon}
\frac{\brakett{m_{1} m_{2}}}{P(m_{2} i)+\epsilon}
\dots
\frac{\brakett{m_{k-2} m_{k-1}}}{P(m_{k-1} i)+\epsilon}
\frac{\brakett{m_{k-1} j}}{P(ji)+\epsilon}.
\end{multline}

The perturbation series for S-matrix elements can be derived from the 
perturbation series above for  $\lim_{T\to\infty}\brakettt{v_i}{\hat{U}_\epsilon (0,-T)}{v_j}$ 
and $\lim_{T\to\infty}\brakettt{v_i}{\hat{U}_\epsilon (0,T)^\dagger}{v_j}$. In the derivation one uses, where necessary, 
the identity $I=\int \intd i\, \ket{v_i}\bra{v_i}$.

If one wants to obtain Taylor series in powers of a coupling constant $g$ appearing in $H$, then one has to take into consideration that generally $H^\ren$, $\ket{v_i}$ and $E_i$ also depend on $g$, therefore the individual terms of a series obtained for a quantity of interest have to be 
expanded further 
into a series in powers of $g$, 
and terms containing the same power of $g$ have to be collected.

\vspace{3mm}
6.) {\sl Lippmann--Schwinger equations}\\
It follows from (\ref{eq.t1}) and (\ref{eq.t12}) that the vectors $\lim_{T\to\infty}\hat{U}_\epsilon (0,-T)\ket{v_j}$ and\\
$\lim_{T\to\infty}\hat{U}_\epsilon (0,T)\ket{v_j}$ satisfy the Lippmann--Schwinger equations
\begin{equation}
\label{eq.ls1}
\lim_{T\to\infty}\hat{U}_\epsilon (0,-T)\ket{v_j}  =  \ket{v_j}+
\frac{-\ii}{\ii (H^{\ren}-E_j)  +\epsilon}
(H-H^\ren)
\lim_{T\to\infty}\hat{U}_\epsilon (0,-T)\ket{v_j},
\end{equation}
\begin{equation}
\lim_{T\to\infty}\hat{U}_\epsilon (0,T)\ket{v_j} =  \ket{v_j}+
\frac{-\ii }{\ii (H^\ren-E_j)  -\epsilon}
(H-H^\ren)
\lim_{T\to\infty}\hat{U}_\epsilon (0,T)\ket{v_j}.
\label{eq.ls2}
\end{equation}
Equations (\ref{eq.t1}) and (\ref{eq.t12}) can also be derived from (\ref{eq.ls1}) and (\ref{eq.ls2}) by iteration. Note that 
$\epsilon$ is a positive number in these equations.

\vspace{3mm}
7.) {\sl Long-range potentials}\\
Long-range potentials, like the Coulomb potential, are well known to require special treatment,
i.e.\ 
(\ref{eq.2in}) and (\ref{eq.2out}) cannot be applied straightforwardly in such cases. 
This situation is not changed by switching to
(\ref{eq.a3}) and (\ref{eq.a4}). One of the possible approaches to handling such potentials
is to introduce a shielding, i.e.\ to approximate a long-range potential by short-range potentials, as described in detail in \cite{Taylor}.

\section{Adiabatic switching}
\label{sec.ad}

Another possible generalization of (\ref{eq.2in}) and (\ref{eq.2out}), that we proposed in a slightly different form in \cite{T}, is the following:
\begin{eqnarray}
\label{eq.7a}
\ket{v,in} & = & \lim_{\epsilon\to +0} \lim_{T\to\infty} 
\tilde{U}_\epsilon(0,-T)\Pi_{\epsilon}(-T)\ket{v},\\
\label{eq.7b}
\ket{w,out} & = & \lim_{\epsilon\to +0} \lim_{T\to\infty} 
\tilde{U}_\epsilon(0,T)\Pi_{\epsilon}(T)\ket{w},
\end{eqnarray}
where
\begin{equation}
\label{eq.Ue}
\tilde{U}_\epsilon(t_2,t_1)
=\mathrm{T}\exp\left[-\ii\int_{t_1}^{t_2}\tilde{H}_{I,\epsilon}(t)\, \intd t\right],
\end{equation}
\beq
\label{eq.pt1}
\tilde{H}_{I,\epsilon}(t)=e^{-\epsilon |t|} e^{\ii H^\ren t}(H-H^\ren) e^{-\ii H^\ren t},
\eeq
and $\mathrm{T}$ denotes time ordering in (\ref{eq.Ue}).
$\Pi_{\epsilon}(T)$ is a unitary operator that is diagonal with respect to a suitable complete orthonormal set of eigenvectors of $H^\ren$, 
and its action on a vector $\ket{v}$ belonging to this set is given by 
\beq
\label{eq.phase}
\Pi_{\epsilon}(T)\ket{v}=
\frac{\sqrt{\brakettt{v}{\tilde{U}_\epsilon(T,0)}{v}\brakettt{v}{\tilde{U}_\epsilon(0,T)}{v}}}{\brakettt{v}{\tilde{U}_\epsilon(0,T)}{v}}\ket{v}.
\eeq

In these formulas an adiabatic switching is applied instead of the abelian limit. 
Equation (\ref{eq.pt1}) implies the replacement of $H$ by the time-dependent Hamiltonian operator 
$\tilde{H}_\epsilon(t)=H^\ren + e^{-\epsilon|t|}(H-H^\ren)$, for which $\lim_{t\to\pm \infty} \tilde{H}_\epsilon (t)= H^\ren$. 
The time-evolution operator 
$\tilde{U}_\epsilon(t_1,t_2)$ is unitary for any $t_1$, $t_2$ and $\epsilon$, therefore it is not necessary 
to include normalization constants in (\ref{eq.7a}) and (\ref{eq.7b}).  The phase operator $\Pi_{\epsilon}(T)$ is 
generally needed, however, to make the $\epsilon\to 0$ limit convergent. 
It does not affect the unitarity of the S-matrix.  
$\tilde{U}_\epsilon(t_1,t_2)$  has the properties $\tilde{U}_\epsilon(t_1,t_2)\tilde{U}_\epsilon(t_2,t_3)=\tilde{U}_\epsilon(t_1,t_3)$
and $\tilde{U}_\epsilon(t_1,t_2)^{-1}=\tilde{U}_\epsilon(t_2,t_1)$.

Adiabatic switching also appears in the literature on scattering theory; in particular, it is known 
that for potential scattering (\ref{eq.7a}) and (\ref{eq.7b}), without the phase operator and with $H^\ren=H_0$, 
where $H_0$ denotes the Hamiltonian operator obtained from $H$ by switching off the interaction,   
yield the same \iin and \oout states as (\ref{eq.2in}) and (\ref{eq.2out}) \cite{Dollard}. 
A similar result was obtained in \cite{T} in perturbation theory.

If $H^\ren$ (and $H$) has a finite, discrete, and nondegenerate spectrum, then (\ref{eq.7a}) is essentially the Gell-Mann--Low formula \cite{GML,FW}, 
which produces eigenvectors of $H$ from eigenvectors of $H^\ren$. This
implies that if also $H_0$ has a nondegenerate spectrum, then $H^\ren=H_0$ can be taken; 
the choice of $H^\ren$ is not restricted.
In particular, it is not necessary 
in (\ref{eq.7a}) and (\ref{eq.7b}) that
$H$ and $H^\ren$ have the same eigenvalues. The phase operator, however, is necessary for the existence of the $\epsilon\to 0$ limit. 

In the case of more general Hamiltonian operators, (\ref{eq.7a}) and  (\ref{eq.7b}) can be regarded as generalizations of the Gell-Mann--Low formula, 
and it is then natural to conjecture that $H^\ren=H_0$, where $H_0$ is defined, as above, 
to be the Hamiltonian operator obtained from $H$ 
by switching off the interaction, 
will be a suitable choice for a large class of models that describe nontrivial scattering processes. 
Perturbative calculations in particular models support this conjecture (see \cite{T} and section \ref{sec.ex}). 

The properties that $H^\ren$ can be taken to be $H_0$ and $\tilde{U}_\epsilon(0,-T)$ and $\tilde{U}_\epsilon(0,T)$ 
are unitary, in contrast with $\hat{U}_\epsilon(0,-T)$ and $\hat{U}_\epsilon(0,T)$,
are appealing features of the formulas (\ref{eq.7a}) and  (\ref{eq.7b}) in comparison with (\ref{eq.a3}) and (\ref{eq.a4}). 
On the other hand, the adiabatic switching is more difficult to handle mathematically than the abelian average.

In view of section \ref{sec.ab} and the result of \cite{Dollard} mentioned above, it is an interesting question whether the phase operator can be omitted if $H^\ren$ has the same spectrum as $H$. 
By perturbative calculations applied to Hamiltonian operators with finite discrete spectra, 
and also to 
the examples discussed in section \ref{sec.ex}, 
one finds that the answer is that this factor cannot be omitted in general. 
One finds the same result by numerical calculations with $2\times 2$ matrices. The analytical results of \cite{B3}, which apply to $2\times 2$ matrices, 
can also be used to show that in general the $\epsilon\to 0$ limit of 
$\lim_{T\to\infty} \tilde{U}_\epsilon(0,-T)\ket{v}$ or 
$\lim_{T\to\infty} \tilde{U}_\epsilon(0,T)\ket{w}$
does not exist if  $H^\ren$ has the same spectrum as $H$.

For perturbative calculations the following formulas can be used.
Let $\{ \ket{v_i}\}$ be an orthonormal basis consisting of eigenvectors of $H^\ren$, and let us denote by $\tilde{E}_i$ the eigenvalues of $H^\ren$ on these vectors.  
From the Dyson series one obtains the following perturbation series  for $\lim_{T\to\infty}\brakettt{v_i}{\tilde{U}_\epsilon (0,-T)}{v_j}$:
\begin{multline}
\label{eq.t2}
\lim_{T\to
  \infty}\brakettt{v_i}{\tilde{U}_\epsilon(0,-T)}{v_j}=
\braket{v_i}{v_j}   -\ii  \frac{\brakett{i j}}{P(i j)+\epsilon}\\ 
+\sum_{k=2}^\infty \left(-\ii \right)^k 
\int \intd m_1 \intd m_2 \dots \intd m_{k-1}
\frac{\brakett{i m_{k-1}}}{P(i j)+k\epsilon}
\frac{\brakett{m_{k-1} m_{k-2}}}{P(m_{k-1} j)+(k-1)\epsilon}
\dots
\frac{\brakett{m_2 m_1}}{P(m_2 j)+2\epsilon}
\frac{\brakett{m_1 j}}{P(m_1 j)+\epsilon},
\end{multline} 
where the notation $P(ij)=\ii (\tilde{E}_i-\tilde{E}_j)$ and $\brakett{ij}= \brakettt{v_i}{H-H^\ren}{v_j}$ is used.

The main difference between this formula and (\ref{eq.t1}) is in the 
coefficients of $\epsilon$ in the denominators. This shows that in general it is essential to carefully take into consideration  
the precise value of these coefficients in order to obtain correct results. 
This point needs to be emphasized because it is usually  not mentioned in the literature. 
Nevertheless, there are also several instances in
calculations when the precise values of these coefficients are not important.

For  $\lim_{T\to\infty}\brakettt{v_i}{\tilde{U}_\epsilon (T,0)}{v_j}$ we have 
\begin{multline}
\label{eq.t22}
\lim_{T\to
  \infty}\brakettt{v_i}{\tilde{U}_\epsilon(T,0)}{v_j}=
\braket{v_i}{v_j} -\ii  \frac{\brakett{i j}}{P(ji)+\epsilon}\\
+\sum_{k=2}^\infty \left(-\ii\right)^k
\int \intd m_1 \intd m_2 \dots \intd m_{k-1}
\frac{\brakett{i m_{1}}}{P(m_1 i)+\epsilon}
\frac{\brakett{m_{1} m_{2}}}{P(m_{2} i)+2\epsilon}
\dots
\frac{\brakett{m_{k-2} m_{k-1}}}{P(m_{k-1} i)+(k-1)\epsilon}
\frac{\brakett{m_{k-1} j}}{P(ji)+k\epsilon}.
\end{multline}

Due to the differences between (\ref{eq.t2}), (\ref{eq.t22}) and (\ref{eq.t1}), (\ref{eq.t12}) 
it is not apparent how equations analogous to 
(\ref{eq.ls1}) and  (\ref{eq.ls2}) could be found for $\lim_{T\to\infty}\tilde{U}_\epsilon (0,-T)\ket{v_j}$ 
and $\lim_{T\to\infty}\tilde{U}_\epsilon (0,T)\ket{v_j}$.

Although not in the context of scattering theory,
adiabatic switching and the Gell-Mann--Low formula have been subject of active research recently; 
see e.g.\  \cite{B3,B1,B2,Molinari,AE, W1,W2} and further references therein.

\section{Examples}
\label{sec.ex}

In this section an application of (\ref{eq.a3}), (\ref{eq.a4}) and (\ref{eq.7a}), (\ref{eq.7b})
in two quantum field theoretical models is presented. Perturbation theory based on (\ref{eq.t1}), (\ref{eq.t12})
and (\ref{eq.t2}), (\ref{eq.t22}) is used
to calculate S-matrix elements and energies of \iin and \oout states.

The first model describes the scattering of a massive relativistic particle (a real scalar boson) on a defect in $1+1$ spacetime dimensions. The defect is localized at $x=0$. The Hamiltonian operator without interaction is
\beq
H_0=\frac{1}{2}\int_{-\infty}^\infty \intd x\ :(\partial_t\phi)^2+(\partial_x\phi)^2+m_0^2\phi^2 :\ , 
\eeq 
where $\phi$ is the boson field, 
satisfying the commutation relation  $[ \phi(x,t),\partial_t \phi(x',t) ] = \ii\delta(x-x')$. 
The interaction term is 
\beq
H_{\mathrm{int}}=:\phi(0,0)^2:\ ,
\eeq
and $H=H_0+gH_{\mathrm{int}}$. 
In this model the interaction term breaks translation and Lorentz symmetry.

The second example is the $\phi^4$ model in $3+1$ dimensions. In this model    
the Hamiltonian operator in the absence of interaction is that of the free scalar boson of mass $m_0$:
\beq
\label{eq.p40}
H_0=\frac{1}{2}\int \intd^3\xx\ :(\partial_t\phi)^2+(\partial_\xx\phi)^2+m_0^2\phi^2 :\ .
\eeq
The interaction term $H_{\mathrm{int}}$ is 
\beq
\label{eq.p4i}
H_{\mathrm{int}}=\int \intd^3\xx\ :\phi^4 : \ ,
\eeq
and $H=H_0+gH_{\mathrm{int}}$.
The field $\phi$ satisfies the commutation relation  $[ \phi(\xx,t),\partial_t \phi(\xx',t) ] = \ii\delta^3(\xx-\xx')$.

In the following, the vacuum--vacuum S-matrix element, the one-particle S-matrix elements, 
in the second example the two-particle S-matrix elements, and the energy of the \iin and \oout states are discussed.
Perturbation theory is applied to second order in $g$ in the first example and to third order in the second example.
The details of the calculations are not presented, since they are rather lengthy and can be considered straightforward.
We note that the application of (\ref{eq.7a}) and (\ref{eq.7b}) in these examples was discussed in \cite{T} as well.
\\

\noindent
{\sl First example: scattering of a scalar particle on a defect}\\
We now turn to the discussion of the application of (\ref{eq.a3}) and (\ref{eq.a4}), in which the abelian average is used, to  
the first example.  We take  $H^\ren$ to be of the form 
\beq
\label{eq.hr}
H^\ren=H_0+\Delta E \cdot I,
\eeq
where $\Delta E$ is a constant that is still to be determined, and $I$ is the unit operator.
In this case
$H_0$ and $H^\ren$ have the same eigenvectors, and one uses as basis vectors the 
vacuum state and the multiparticle plane-wave eigenstates of $H_0$.
The vacuum state $\ket{\Omega}$ is normalized as $\braket{\Omega}{\Omega}=1$, and 
the one- and two-particle states 
$\ket{k}$ and $\ket{k_1 k_2}$ as
$\braket{k_1}{k_2}=\delta(k_1-k_2)$, $\braket{k_3 k_4}{k_1 k_2}=\delta(k_1-k_3)\delta(k_2-k_4)+
\delta(k_1-k_4)\delta(k_2-k_3)$, etc.

The ansatz (\ref{eq.hr}) for $H^\ren$ can be regarded as a guess, but 
one can also infer that this should be the choice for $H^\ren$ from the results of the application of (\ref{eq.7a}) and (\ref{eq.7b}) (in which the adiabatic switching is used). In that case one takes $H^\ren=H_0$, and then one can calculate the eigenvalues that $H$ has on the \iin and \oout states. 
In this way one obtains that the energies of the \iin and \oout states are shifted by a common constant with respect to their values at $g=0$.

The coefficients in the Taylor series for $\Delta E$ are denoted as 
$\Delta E= g\Delta E^{(1)}+g^2 \Delta E^{(2)}+\dots$. 
To second order, one finds that the requirement that the $\epsilon\to +0$ limit of \\
$\lim_{T\to\infty}
\brakettt{\Omega}{\hat{U}_{\epsilon}(0,T)^\dagger\hat{U}_{\epsilon}(0,-T)}{\Omega}$ 
be convergent determines  $\Delta E^{(1)}$ and  $\Delta E^{(2)}$ uniquely, and one obtains  
$\Delta E^{(1)}=0$ and 
\begin{equation}
\Delta E^{(2)}
=-\int \intd k_A \intd k_B\, \frac{1}{32\pi^2}\frac{1}{\omega_A\omega_B(\omega_A+\omega_B)},
\label{eq.q1}
\end{equation}
where the notation $\omega=\sqrt{m_0^2+k^2}$ has been used.
For $Z_{\Omega,in}$  and $Z_{\Omega,out}$, one then obtains
\begin{eqnarray}
Z_{\Omega,in}= Z_{\Omega,out}= 1-g^2\int \intd k_A \intd k_B\, \frac{1}{32\pi^2}
\frac{1}{\omega_A\omega_B(\omega_A+\omega_B)^2} + O(g^3).
\label{eq.q2}
\end{eqnarray}
The final result for the vacuum--vacuum S-matrix element to second order in $g$ is $S_{\Omega;\Omega}=\braket{\Omega,out}{\Omega,in}=1+O(g^3)$, as one expects.
Equation (\ref{eq.q2}) is also consistent with $Z_{\Omega,in}\le 1$, since the integrand is positive.

Regarding one-particle states (with $k\ne 0$), one finds that if $\Delta E^{(1)}$ and  $\Delta E^{(2)}$ are chosen as above, then
the $\epsilon\to +0$  limit of 
$\lim_{T\to \infty}\brakettt{k_2}{\hat{U}_{\epsilon}(0,T)^\dagger\hat{U}_{\epsilon}(0,-T)}{k_1}$ 
is convergent.
One obtains that $\lim_{T\to \infty}
\brakettt{k_2}{\hat{U}_{\epsilon}(0,-T)^\dagger\hat{U}_{\epsilon}(0,-T)}{k_1}=
Z_{\Omega,in}\delta(k_1-k_2)$, hence $Z_{k,in}=Z_{\Omega,in}$. One also finds that $Z_{k,out}=Z_{k,in}$, to second order in $g$.

We conjecture that to all orders in $g$ there exists a unique value of $\Delta E$, such that the $\epsilon\to +0$  limit of 
$\lim_{T\to \infty}\brakettt{v_i}{\hat{U}_{\epsilon}(0,T)^\dagger\hat{U}_{\epsilon}(0,-T)}{v_j}$ is convergent 
for any plane-wave basis vectors $\ket{v_i}$ and $\ket{v_j}$ (not containing any particle with $k=0$), 
and that 
the $Z$ factor is the same for all such states, to all orders in $g$.

We find that to second order the S-matrix element $S_{k_1; k_2}$ is 
\beq
\label{eq.str}
S_{k_1;k_2}=T(k_2)\delta(k_1-k_2)+R(k_2)\delta(k_1+k_2),
\eeq
where
\beq
T(k_2)=1-g\frac{\ii }{2|k_2|}-g^2\frac{1}{4 k_2^2} +O(g^3)\ , \qquad
R(k_2)=-g\frac{\ii}{2|k_2|}-g^2\frac{1}{4k_2^2}+O(g^3).
\eeq 
An exact result for $S_{k_1; k_2}$ to all orders was obtained  in a different framework in \cite{DMS1}. It 
also takes the form (\ref{eq.str}), and $T(k_2)$ and $R(k_2)$ are\footnote{This result was quoted with some wrong signs in \cite{T}.} 
\beq
T(k_2)=\frac{\ii|k_2|}{\ii|k_2|-g/2}\ ,\qquad
R(k_2)=\frac{g/2}{\ii|k_2|-g/2}\ .
\eeq

\medskip
The application of equations (\ref{eq.7a}) and (\ref{eq.7b}), in which the adiabatic switching 
is used,
with $H^\ren=H_0$, yields the same  results for $S_{\Omega;\Omega}$ and $S_{k_1; k_2}$ as above.
The phase factors in (\ref{eq.7a}) and (\ref{eq.7b}) are necessary for obtaining convergent results as $\epsilon\to 0$. 
The states
$\ket{\Omega,in}=\ket{\Omega,out}$, $\ket{k,in}$ and $\ket{k,out}$ are eigenvectors of $H$, therefore their 
energies can be calculated from the eigenvalue equation.
One finds that the energies of these states are all shifted by a common constant 
with respect to their value at $g=0$; the
value of this constant equals to $\Delta E$ above (at least to second order in $g$).\\

\noindent
{\sl Second example: $\phi^4$ model}\\
Continuing with the second example,
it is well known that in perturbation theory the $\phi^4$ model in $3+1$ dimensions contains ultraviolet divergences, 
which can for example be handled by introducing a cutoff. 
For simplicity we shall not introduce such a cutoff, therefore our results will be formal expressions, which, however, 
is sufficient for our present purpose. One could also consider the $\phi^4$ model in $1+1$ dimensions; 
in that case the calculations and the results are very similar to those in $3+1$ dimensions, but ultraviolet divergences do not occur. 
It is also well known that there are divergences in the model associated with translation invariance and the infiniteness of the volume of space. 
In perturbation theory these divergences are related to vacuum bubble diagrams (see e.g.\ \cite{PS}, chapter 4). 
A rigorous treatment of these divergences could be given by introducing a suitable regularization, e.g.\ by considering the system in a finite volume. In this case the space volume has to be increased simultaneously with $T$ 
(that appears in (\ref{eq.a3}), (\ref{eq.a4}) and in (\ref{eq.7a}), (\ref{eq.7b})). 
Nevertheless, it is also known that the contributions of the vacuum bubble diagrams cancel out from correlation functions and from S-matrix elements in the standard formalism, although the demonstration of this result is usually done in a formal manner, i.e.\ without doing a rigorous regularization 
(see e.g.\ chapter 17 of \cite{BD} and chapter 6 of \cite{IZ}). 
In the present paper we also restrict ourselves to a formal treatment of the divergences corresponding to vacuum bubble diagrams, mainly because a rigorous treatment would be much more complicated, 
and because we do not expect that it would change the results. 
For further details on the difficulties of defining 
Poincar\'e-symmetric quantum field theoretical models, we refer the reader e.g.\ to 
section 4-1-1 of \cite{IZ} and to \cite{Haag}. We also note that one of our reasons for including the first example in this paper is that although
it is a quantum field theory, it does not have divergences in perturbation theory that correspond to vacuum bubble diagrams, and it is also free of ultraviolet divergences.

We consider the application of (\ref{eq.a3}) and (\ref{eq.a4}) first. 
If we kept $H=H_0+gH_{\mathrm{int}}$ as it is specified at the beginning of this section, with $H_0$ and $H_{\mathrm{int}}$ given by 
(\ref{eq.p40}) and (\ref{eq.p4i}),
then the mass of the particles described by $H$ would be different from $m_0$, therefore $H^\ren$ would depend on $g$, and the relation between 
$H_0$ and $H^\ren$ would be more complicated than in the previous example (i.e.\  $H^\ren$ would not be of the form $\Delta E \cdot I + H_0$).
This would complicate the application of (\ref{eq.t1}) and (\ref{eq.t12}), since these formulas are given in terms of eigenvalues of $H^\ren$ and 
matrix elements with respect to eigenstates of $H^\ren$, whereas the known quantities, in terms of which the model is defined, are 
the eigenvalues of $H_0$ and the matrix elements of $H_{\mathrm{int}}$ with respect to the 
plane-wave and vacuum eigenstates of $H_0$.
For this reason     
we modify $H$ by adding two terms, so that the modified total Hamiltonian operator that we actually consider is
\beq
\label{eq.hv}
H'=H_0+g \int \intd^3\xx\ :\phi^4 :  \  -\,\Delta E\cdot I - \frac{1}{2}\delta m^2\int \intd^3\xx\  :\phi^2 : \ .
\eeq
The $\Delta E$ and $\delta m^2$ parameters introduced here should be chosen in such a way 
that the ground-state energy and the particle mass corresponding to
$H'$ be $0$ and $m_0$, respectively, so that $H^\ren=H_0$ can be taken. The asymptotic states from which 
the scattering states are produced by (\ref{eq.a3}) and (\ref{eq.a4}) are then 
the vacuum state and the multiparticle plane-wave eigenstates of $H_0$, and the matrix elements and eigenvalues entering the formulas
(\ref{eq.t1}), (\ref{eq.t12}) are known, thus the difficulty described above is avoided. 
Nevertheless,  $\Delta E$ and $\delta m^2$ are initially unknown constants that depend on $g$ and have to be determined. 
We note that the kind of modification of the model that we have done is common as a part of renormalization.    

The $-\Delta E\cdot I$ term does not have much physical significance, as it merely shifts all energies by a common constant. 
The $-\frac{1}{2}\delta m^2\int \intd^3\xx\  :\phi^2 :$ term serves to readjust the mass of the particles described by $H'$ to the value $m_0$, 
which is 
the physical value in the present approach. 
The coefficients in the Taylor series of  $\Delta E$ and $\delta m^2$  are denoted as 
$\Delta E= g\Delta E^{(1)}+g^2 \Delta E^{(2)}+\dots$ and $\delta m^2= g(\delta m^2)^{(1)}+g^2 (\delta m^2)^{(2)}+\dots$.

To second order, $\Delta E^{(1)}$ and  $\Delta E^{(2)}$ are determined in a similar way as in the first example, and one 
obtains  
$\Delta E^{(1)}=0$ and 
\begin{equation}
\Delta E^{(2)}=  -\int \frac{\intd^3 \kk_A \intd^3 \kk_B \intd^3 \kk_C \intd^3 \kk_D}{4!}
\frac{\eta^2}{\omega_A\omega_B\omega_C\omega_D} 
\frac{[\delta^3(\kk_A+\kk_B+\kk_C+\kk_D)]^2}{\omega_A+\omega_B+\omega_C+\omega_D},
\label{eq.b1}
\end{equation}
where
$\eta=4!/4(2\pi)^3$, and the notation $\omega=\sqrt{m_0^2+\kk^2}$ is used.

For $Z_{\Omega,in}$  and $Z_{\Omega,out}$ one obtains
\begin{eqnarray}
Z_{\Omega,in}= Z_{\Omega,out}=  1-
g^2\int \frac{\intd^3 \kk_A \intd^3 \kk_B \intd^3 \kk_C \intd^3 \kk_D}{4!}\frac{\eta^2}{\omega_A\omega_B\omega_C\omega_D} \nonumber \\
\times \frac{[\delta^3(\kk_A+\kk_B+\kk_C+\kk_D)]^2}{(\omega_A+\omega_B+\omega_C+\omega_D)^2} + O(g^3).
\label{eq.zz}
\end{eqnarray}
The integrand on the right-hand side is nonnegative, which is consistent with $Z_{\Omega,in}\le 1$. 
The final result for the vacuum--vacuum S-matrix element 
is $S_{\Omega;\Omega}=\braket{\Omega,out}{\Omega,in}=1+O(g^3)$, in agreement with the expectation.

Regarding $\lim_{T\to \infty}
\brakettt{\kk_2}{\hat{U}_{\epsilon}(0,T)^\dagger\hat{U}_{\epsilon}(0,-T)}{\kk_1}$, we obtain 
from the requirement that the $\epsilon\to +0$  limit of this quantity should be convergent
that $(\delta m^2)^{(1)}=0$  and 
\begin{eqnarray}
(\delta m^2)^{(2)} & = & -2\int \frac{\intd^3 \kk_A \intd^3 \kk_B  }{3!}
\frac{\eta^2}{\omega_A\omega_B\omega_C(\omega_A+\omega_B+\omega_C-m_0)} \nonumber\\
&& -2\int \frac{\intd^3 \kk_A \intd^3 \kk_B  }{3!}
\frac{\eta^2}{\omega_A\omega_B\omega_C(\omega_A+\omega_B+\omega_C+m_0)}, 
\label{eq.b2}
\end{eqnarray}
where $\kk_C=-\kk_A-\kk_B$ in $\omega_C$.

\begin{figure}[t]
\begin{center}
\includegraphics[scale=0.4]{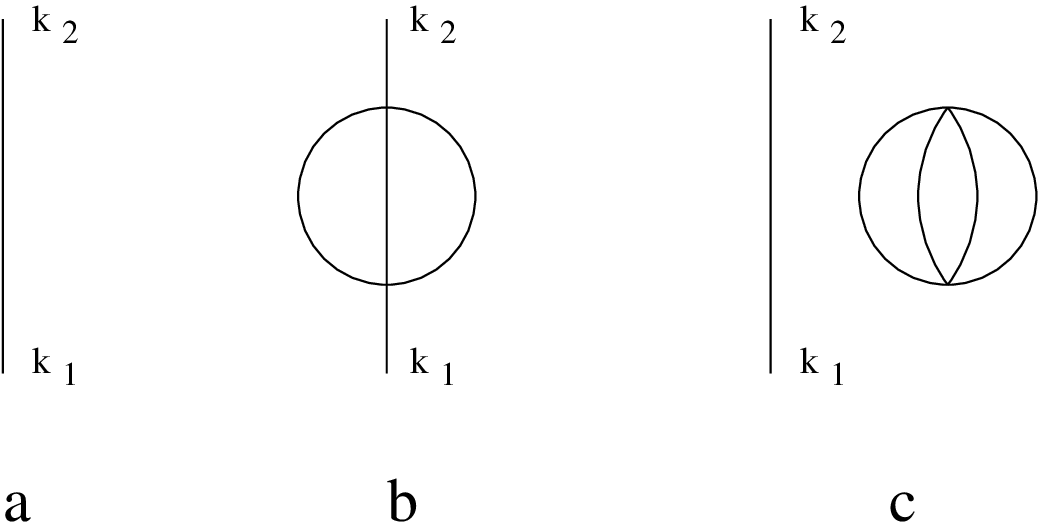}
\end{center}
\caption{\label{op}
{\em Graphs for the one-particle S-matrix element  to second order}}
\end{figure}

\begin{figure}
\begin{center}
\includegraphics[scale=0.4]{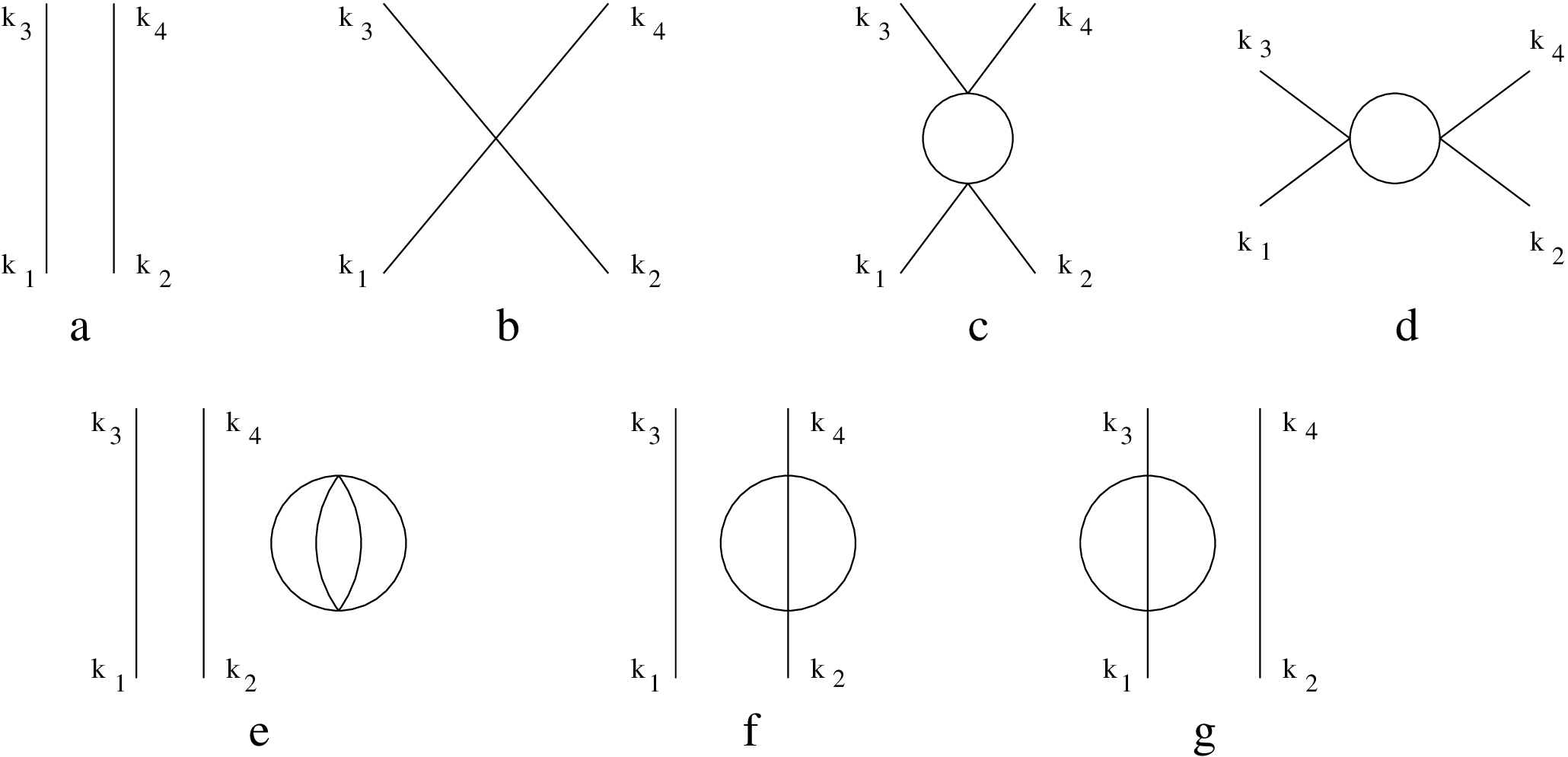}
\end{center}
\caption{\label{tp}
{\em Graphs for the two-particle S-matrix elements  to second order}}
\end{figure}

\begin{figure}
\begin{center}
\includegraphics[
  scale=0.4]{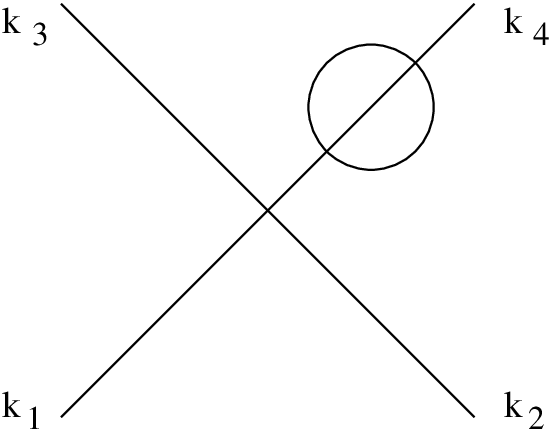}
\end{center}
\caption{\label{rc}
{\em A third-order graph containing a radiative correction on an external line}}
\end{figure}

For  $Z_{\kk,in}$ and $Z_{\kk,out}$ we obtain
\begin{eqnarray}
Z_{\kk,in}\ =\ Z_{\kk,out} & = & Z_{\Omega,in} \nonumber \\
&& - g^2\int \frac{\intd^3 \kk_A \intd^3 \kk_B  }{3!}\frac{\eta^2}{\omega_A\omega_B\omega_C\omega}  
\frac{1}{(\omega_A+\omega_B+\omega_C-\omega)^2} \nonumber \\
&& - g^2\int \frac{\intd^3 \kk_A \intd^3 \kk_B  }{3!}\frac{\eta^2}{\omega_A\omega_B\omega_C\omega} 
\frac{1}{(\omega_A+\omega_B+\omega_C+\omega)^2 } + O(g^3).
\end{eqnarray}
This result is also consistent with $Z_{\kk,in}\le 1$.
The final result for the one-particle S-matrix element is 
$\braket{\kk_2,out}{\kk_1,in}=\delta^3(\kk_2-\kk_1) + O(g^3)$, which is also in agreement with the expectation.
Our result for $\delta m^2$ 
is  equal to the result that can be obtained in the standard field theoretical formalism.

To second order, the contributions to
$\lim_{T\to \infty}
\brakettt{\kk_2}{\hat{U}_{\epsilon}(0,T)^\dagger\hat{U}_{\epsilon}(0,-T)}{\kk_1}$ arising from the interaction term $g \int \intd^3\xx\ :\phi^4 :$ 
correspond to the graphs shown in figure \ref{op}. (Graphs can be associated with various terms in perturbation theory in a similar way as in the standard formalism. Detailed graph rules are not described here, since they are not needed in this paper.) 
The contributions corresponding to \ref{op}.b and 
\ref{op}.c diverge as $1/\epsilon$ in the $\epsilon\to +0$ limit. These divergences are canceled out by further contributions arising from the terms 
$-\Delta E\cdot I$ and\\ $-\frac{1}{2}\delta m^2\int \intd^3\xx\  :\phi^2 :$ in $H'$.

Considering $\lim_{T\to \infty}
\brakettt{\kk_3,\kk_4}{\hat{U}_{\epsilon}(0,T)^\dagger\hat{U}_{\epsilon}(0,-T)}{\kk_1,\kk_2}$,
we find that its $\epsilon\to 0$ limit is finite if  $\Delta E^{(1)}$,  $\Delta E^{(2)}$,  
$(\delta m^2)^{(1)}$ and $(\delta m^2)^{(2)}$ are chosen as above.

For  $Z_{\kk_1\kk_2,in}$ and $Z_{\kk_1\kk_2,out}$ one finds  
\beq
Z_{\kk_1\kk_2,in}=Z_{\kk_1\kk_2,out}
\eeq
and 
\beq
\frac{Z_{\kk_1\kk_2,in}}{Z_{\Omega,in}}=
\frac{Z_{\kk_1,in}}{Z_{\Omega,in}}
\frac{Z_{\kk_2,in}}{Z_{\Omega,in}}.
\eeq
It is natural to expect that these relations hold to all orders of perturbation theory, and that similar relations hold for states containing 
an arbitrary number of particles.

The diagrams corresponding to the contributions to
$\lim_{T\to \infty}
\brakettt{\kk_3,\kk_4}{\hat{U}_{\epsilon}(0,T)^\dagger\hat{U}_{\epsilon}(0,-T)}{\kk_1,\kk_2}$ arising from $g \int \intd^3\xx\ :\phi^4 :$ in $H'$
to second order
are those that are shown in figure \ref{tp}, and those that are obtained from \ref{tp}.a, \ref{tp}.d, \ref{tp}.e, \ref{tp}.f and \ref{tp}.g
by interchanging $\kk_3$ and $\kk_4$.
The terms that are divergent in the $\epsilon\to +0$ limit correspond to the graphs \ref{tp}.e, \ref{tp}.f and \ref{tp}.g 
and to those that are obtained from \ref{tp}.e, \ref{tp}.f and \ref{tp}.g by interchanging $\kk_3$ and $\kk_4$. The rate of divergence of these terms is $1/\epsilon$. Further terms arising from the counterterms 
$-\Delta E\cdot I$ and $-\frac{1}{2}\delta m^2\int \intd^3\xx\  :\phi^2 :$ cancel out these divergences. 

The final result for the two-particle S-matrix contains only the terms that correspond to the graphs \ref{tp}.a, \ref{tp}.b, \ref{tp}.c and \ref{tp}.d 
(and to those obtained from these graphs by interchanging $\kk_3$ and $\kk_4$).

To first order the two-particle S-matrix reads as
\begin{eqnarray}
S_{\kk_3\kk_4;\kk_1\kk_2} & = & \delta^3(\kk_1-\kk_3)\delta^3(\kk_2-\kk_4)+\delta^3(\kk_1-\kk_4)\delta^3(\kk_2-\kk_3) \nonumber \\
& & -g 2\pi\ii  \delta(\omega_1+\omega_2-\omega_3-\omega_4) \delta^3(\kk_1+\kk_2-\kk_3-\kk_4)  \frac{\eta}{\sqrt{\omega_1\omega_2\omega_3\omega_4}} + O(g^2).
\end{eqnarray}

At third order, we focus on those
contributions to 
$\lim_{T\to \infty}
\brakettt{\kk_3,\kk_4}{\hat{U}_{\epsilon}(0,T)^\dagger\hat{U}_{\epsilon}(0,-T)}{\kk_1,\kk_2}$ yielded by
the interaction term $g\int \intd^3\xx\ :\phi^4 :$ 
which correspond to graphs that contain a radiative correction on an external line. 
One of these graphs  is
shown in figure \ref{rc}, and there are three other similar ones in which the correction is on the $\kk_1$, $\kk_2$ or $\kk_3$ leg.
These contributions diverge
as $1/\epsilon$ in the  $\epsilon\to +0$ limit. This divergence is canceled out by further terms arising from 
the counterterm 
$-\frac{1}{2}\delta m^2\int \intd^3\xx\  :\phi^2 :$ in $H'$,  with   $(\delta m^2)^{(1)}$ and $(\delta m^2)^{(2)}$ as above. 
There are also third-order contributions to $S_{\kk_3\kk_4;\kk_1\kk_2}$ from the product of the second-order parts of 
$1/\sqrt{Z_{\kk_1\kk_2}}$ and  $1/\sqrt{Z_{\kk_3\kk_4}}$ and the first-order part of 
$\lim_{\epsilon\to+0}\lim_{T\to \infty}
\brakettt{\kk_3,\kk_4}{\hat{U}_{\epsilon}(0,T)^\dagger\hat{U}_{\epsilon}(0,-T)}{\kk_1,\kk_2}$, 
which is $-g 2\pi\ii  \delta(\omega_1+\omega_2-\omega_3-\omega_4) \delta^3(\kk_1+\kk_2-\kk_3-\kk_4)  \frac{\eta}{\sqrt{\omega_1\omega_2\omega_3\omega_4}}$ 
(the graph  corresponding to this factor is depicted in figure \ref{tp}.b). The sum of these terms 
plus the first-order part is given by 
\beq
\label{eq.fo}
-g 2\pi\ii  \delta(\omega_1+\omega_2-\omega_3-\omega_4) \delta^3(\kk_1+\kk_2-\kk_3-\kk_4)  \frac{\eta}{\sqrt{\omega_1\omega_2\omega_3\omega_4}}
\sqrt{Z_\phi}^4 + O(g^4),
\eeq
where $Z_\phi$ is the constant
\begin{eqnarray}
Z_\phi & = & 1- g^2\int \frac{\intd^3 \kk_A \intd^3 \kk_B  }{3!}\frac{\eta^2}{\omega_A\omega_B\omega_C m_0}  
\frac{1}{(\omega_A+\omega_B+\omega_C-m_0)^2} \nonumber \\
&& + g^2\int \frac{\intd^3 \kk_A \intd^3 \kk_B  }{3!}\frac{\eta^2}{\omega_A\omega_B\omega_C m_0} 
\frac{1}{(\omega_A+\omega_B+\omega_C+m_0)^2 } +O(g^3),
\label{eq.zphi}
\end{eqnarray}
with $\kk_C=-\kk_A-\kk_B$ in $\omega_C$. 
This $Z_\phi$ is equal to the field-strength renormalization constant of the standard formalism.
Although $Z_\phi$ has some similarity to $Z_{\kk}/Z_\Omega$, they are different quantities; in particular the latter is not independent of $\kk$.

The expression (\ref{eq.fo}) is the same as the one obtained using the standard formalism. 
Of course, there are further contributions to $S_{\kk_3\kk_4;\kk_1\kk_2}$ at order $g^3$, 
which correspond to connected graphs that contain three vertex points 
(of order four) and do not contain corrections on external lines. 
Denoting these contributions, together with those at order $g^2$ which coorrespond to connected graphs that contain two vertex points of order four, by ..., to third order we have
\begin{eqnarray}
S_{\kk_3\kk_4;\kk_1\kk_2} 
& = & \delta^3(\kk_1-\kk_3)\delta^3(\kk_2-\kk_4)+\delta^3(\kk_1-\kk_4)\delta^3(\kk_2-\kk_3)\nonumber\\
&& -g 2\pi\ii  \delta(\omega_1+\omega_2-\omega_3-\omega_4) \delta^3(\kk_1+\kk_2-\kk_3-\kk_4)  \frac{\eta}{\sqrt{\omega_1\omega_2\omega_3\omega_4}} \sqrt{Z_\phi}^4 \nonumber\\
&& + \dots +O(g^4).
\label{eq.v0}
\end{eqnarray}

\medskip
In the case of the formulas (\ref{eq.7a}) and (\ref{eq.7b}), in which the adiabatic switching is used, we take $H^\ren=H_0$, and we also leave $H$ unchanged.
We make the choice $H^\ren=H_0$ because $H_0$ is the operator obtained from $H$ by switching off the interaction, i.e.\ by setting $g$ to zero, 
thus this is the choice that one would try first. 
Another reason is that a more complicated $H^\ren$ would entail the same difficulties as 
described above at the beginning of the discussion of the application of (\ref{eq.a3}) and (\ref{eq.a4}). 
In addition, by doing calculations with the choice 
$H^\ren=H_0$ (and with the original $H$) one can
illustrate the feature of (\ref{eq.7a}) and (\ref{eq.7b}) that they do not require  
the ground-state energy and particle mass corresponding to $H^\ren$ and $H$ to be the same.

Regarding the perturbation series for the quantities  
$\lim_{T\to \infty}
\brakettt{\Omega}{\tilde{U}_{\epsilon}(0,T)^\dagger\tilde{U}_{\epsilon}(0,-T)}{\Omega}$,\\
$\lim_{T\to \infty}
\brakettt{\kk_2}{\tilde{U}_{\epsilon}(0,T)^\dagger\tilde{U}_{\epsilon}(0,-T)}{\kk_1}$ and
$\lim_{T\to \infty}
\brakettt{\kk_3,\kk_4}{\tilde{U}_{\epsilon}(0,T)^\dagger\tilde{U}_{\epsilon}(0,-T)}{\kk_1,\kk_2}$
to second or third order in $g$, one finds that the coefficients of the higher than first-order terms 
do not have finite limits as $\epsilon\to 0$, and the divergent parts correspond to graphs that contain disconnected vacuum bubbles or 
radiative corrections on external lines, similarly to the case when the abelian limit is applied.
Nevertheless, the phase factors that are also included in  (\ref{eq.7a}) and (\ref{eq.7b}) cancel out these divergences, 
thus for the 
S-matrix elements $S_{\Omega;\Omega}$, $S_{\kk_1;\kk_2}$ and $S_{\kk_3\kk_4;\kk_1\kk_2}$ finite results are obtained. 

The final results for $S_{\Omega;\Omega}$ and  $S_{\kk_1;\kk_2}$ are the same as above, as one expects; 
$S_{\Omega;\Omega}=1+O(g^3)$ and $S_{\kk_1;\kk_2}=\delta^3(\kk_2-\kk_1)+O(g^3)$.
To third order we obtain for $S_{\kk_3\kk_4;\kk_1\kk_2}$ the result
\begin{eqnarray}
S_{\kk_3\kk_4;\kk_1\kk_2} 
& = & \delta^3(\kk_1-\kk_3)\delta^3(\kk_2-\kk_4)+\delta^3(\kk_1-\kk_4)\delta^3(\kk_2-\kk_3)\nonumber\\
&& -g 2\pi\ii  \delta(\hat{\omega}_1+\hat{\omega}_2-\hat{\omega}_3-\hat{\omega}_4) \delta^3(\kk_1+\kk_2-\kk_3-\kk_4)  \frac{\eta}{\sqrt{\hat{\omega}_1\hat{\omega}_2\hat{\omega}_3\hat{\omega}_4}} \sqrt{Z_\phi}^4 \nonumber\\
&& + \dots +O(g^4),
\label{eq.v}
\end{eqnarray}
where ... stands for further terms corresponding to connected diagrams that contain two or three vertices of order four and that do not contain corrections on external lines. 
These terms are the same as in the case above where the abelian limit is used. 
In (\ref{eq.v}), $\hat{\omega}_1$ denotes $\sqrt{m^2+\kk_1^2}$ (etc), where $m^2=m_0^2+\delta m^2$ and 
the $\delta m^2$ here is equal to the $\delta m^2$ found above in the case of the abelian limit.
$Z_\phi$ is also the same as in (\ref{eq.zphi}).
The differences between the results (\ref{eq.v0}) and  (\ref{eq.v})  obtained
for $S_{\kk_3\kk_4;\kk_1\kk_2}$ using (\ref{eq.7a}), (\ref{eq.7b}) and 
(\ref{eq.a3}), (\ref{eq.a4}), respectively, are
due to the fact that the total Hamiltonians in the two cases are 
different; specifically the particle masses corresponding to them are not the same.  

The energies of $\ket{\Omega,in}$ and $\ket{\kk,in}$ can also be calculated from the eigenvalue equation (as these states are eigenstates of $H$). This 
yields the result that to second order the energy of $\ket{\Omega,in}$ is equal to $\Delta E$, and the square of the mass of the scalar particle is equal to $m_0^2+\delta m^2$.   

For a more direct comparison with the case where the abelian limit is applied, one can take $H'$ given by  (\ref{eq.hv}), (\ref{eq.b1}) and 
(\ref{eq.b2})
as the total Hamiltonian operator, and   
$H^\ren=H_0$. The results for  $S_{\Omega;\Omega}$, $S_{\kk_1;\kk_2}$ and $S_{\kk_3\kk_4;\kk_1\kk_2}$
with these Hamiltonians are then exactly the same as above where the abelian limit is applied. 
The phase factors included in  (\ref{eq.7a}) and (\ref{eq.7b}) are still necessary; 
without these factors one would not obtain convergent (as $\epsilon\to 0$) final results.

\section{Summary and conclusion}
\label{sec.c}

We discussed the problem of the generalization of quantum-mechanical formal scattering theory to a wider
class of models that includes quantum field theories, with the aim of clarifying certain 
elementary aspects that are not treated completely satisfactorily in textbooks. We intended to draw attention to the fact 
that in the case of quantum field theories the most straightforward strong limit that can be applied 
in the standard quantum-mechanical formula for producing the \iin and \oout states is not suitable, 
therefore some more effective limiting procedure is needed, which requires certain modifications of the formulas used in quantum mechanics. 

We studied two possibilities: the application of the abelian limit and of adiabatic switching.
The modifications of the quantum-mechanical formulas required by these two methods are significantly different, which shows that 
quantum-mechanical formal scattering theory can be generalized in at least two nontrivially different ways. 
Neither of the two ways appear to be distinguished.  
 
The first method that we considered, which was the abelian limit, 
does not preserve the unitarity of the time-evolution operator, 
making it necessary to include suitable normalization factors in the formulas for the  \iin and \oout states. 
The orthogonality properties of the \iin and \oout states also need verification. 
In addition, the application of the abelian limit requires that the particle masses and the vacuum energy 
corresponding to the asymptotic Hamiltonian operator that describes the particles far in time from the 
collision be the same as the masses and the vacuum energy 
corresponding to the full Hamiltonian operator. 

The second method, the application of an adiabatic switching, 
preserves the unitarity of the time-evolution operator,
therefore it is not necessary to introduce normalization factors into the formulas for the \iin and \oout states. However, it is necessary to include
suitable compensating phase factors in the formulas to make the phases of the \iin and \oout states convergent when the $\epsilon\to 0$ limit is taken.
It is also not necessary that the particle masses and the vacuum energy corresponding to the asymptotic Hamiltonian operator be the same as 
for the full Hamiltonian operator. 
While a Lippmann--Schwinger equation can be written down in the case when the abelian limit is applied, 
this does not seem to be possible in a straightforward way 
in the case of adiabatic switching.

In order to provide illustration of the features of the two approaches, we 
considered the case when the Hilbert space is finite dimensional,
and we also
studied two specific quantum field theoretical models. 
The case of finite-dimensional Hilbert space is interesting because of its mathematical simplicity and because 
it shows that the features mentioned above are not specific to infinite-dimensional Hilbert spaces or to
quantum field theory.
For the two quantum field theoretical models 
we calculated the vacuum--vacuum, 
one-particle and two-particle S-matrix elements 
and the energy of the \iin and \oout states
to second (and in some cases to third) order
in the coupling constant, using suitably modified old-fashioned perturbation theory. 
The results of these calculations confirm the necessity of the modifications of the 
quantum-mechanical formulas, and 
they are also in agreement with results that can be obtained by other methods,   
indicating that the formulas (\ref{eq.a3}), (\ref{eq.a4}) and  (\ref{eq.7a}), (\ref{eq.7b}) can 
indeed be applied to quantum field theories.
It remains an open problem to extend our results to higher orders of perturbation theory.

Although expected, 
it is remarkable that both  methods considered in this paper, in spite of the differences between them, give equivalent final results, at least for the 
physical quantities above and to the orders taken into account. 
It would be interesting to investigate whether there exist models for which the two methods give different results.  
A further interesting question is whether the S-matrix elements yielded by (\ref{eq.a3}), (\ref{eq.a4}) or 
(\ref{eq.7a}), (\ref{eq.7b}) can depend on the choice 
of $H^\ren$, with fixed total Hamiltonian operator.


\end{document}